\documentclass{article}
\usepackage{spconf,amsmath,graphicx}

\title{AUTHOR GUIDELINES FOR ICASSP 2020 PROCEEDINGS MANUSCRIPTS}
\name{Author(s) Name(s)}%\thanks{Thanks to XYZ agency for funding.}}
\address{Author Affiliation(s)}
\twoauthors
  {Shulai Zhang}
	{Dept. of Electronic Engineering\\
	Shanghai Jiao Tong University\\
	Shanghai, P. R. China}
  {Xiaoli Ma}
	{School of ECE\\
	Georgia Institute of Technology\\
	Atlanta, GA, USA}
\usepackage{cite}
\usepackage{amsmath,amssymb,amsfonts}
\usepackage{url}
\usepackage{caption}
\usepackage{algorithm}
\usepackage{graphicx}
\usepackage{subfigure}
\usepackage{textcomp}
\usepackage{xcolor}

\begin{document}
\title{A General Difficulty Control Algorithm for Proof-of-Work Based Blockchains}
%\author{Shulai Zhang} 
%
%\authorrunning{Shulai Zhang}
%\institute{
%\email{zslzsl1998@sjtu.edu.cn}\\
%}
%
\maketitle              % typeset the header of the contribution
\begin{abstract}
Designing an efficient difficulty control algorithm is an essential problem in Proof-of-Work (PoW) based blockchains because the network hash rate is randomly changing. This paper proposes a general difficulty control algorithm and provides insights for difficulty adjustment rules for PoW based blockchains. The proposed algorithm consists a two-layer neural network. It has low memory cost, meanwhile satisfying the fast-updating and low volatility requirements for difficulty adjustment. Real data from Ethereum are used in the simulations to prove that the proposed algorithm has better performance for the control of the block difficulty.

%\keywords{Difficulty control algorithms  \and Simulation methods.}
\end{abstract}


\section{Introduction}

Blockchain has created a huge wave in manufacturing industry, academic institutions, and financing organizations. The bottomline of a blockchain is to generate a decentralized ledger which consists of a sequence of blocks and each block contains the information of some transactions. The way of generating the blocks depends on the consensus algorithms. Currently, the well-adopted consensus algorithm is Proof-of-Work (PoW) \cite{dwork1992pricing}, \cite{back2002hashcash}, which is used in both Bitcoin\cite{nakamoto2008bitcoin} and Ethereum\cite{wood2014ethereum}.

In a PoW based blockchain, blocks are generated by users who participate in mining. The users are called miners, whose computing power is used to solve a difficult mathematical puzzle for each block. The expected time to find a solution is proportional to the difficulty of the puzzle and depends on the total computing power (hash rate) provided by all users in the network. The miner who finds a solution for the puzzle has the right to publish his newly created block to all other users. The time between the generated timestamps of continuous blocks is called the block produce time.

Since block difficulty determines the block produce time directly, difficulty control is extraordinarily significant for the stability of a PoW based blockchain. A proper difficulty control algorithm is supposed to promise the consistency in generating blocks and promise the block produce time to converge to the desired value no matter how the network hash rate changes to ensure efficient transactions. In \cite{kraft2016difficulty}, an improved difficulty control algorithm for Bitcoin is proposed based on the assumption that the hash rate is exponentially increasing. An algorithm able to reduce an incentive to perform coin-hopping attack for Bitcoin is proposed in \cite{meshkov2017short} and an algorithm that collects hash rate commitments secured by bond from miners is proposed in \cite{bissias2019bonded}. The work \cite{fullmer2018analysis} presents a stochastic model for the block produce time and analyzes the marginal distribution of the block produce time in Bitcoin. However, a general difficulty control algorithm for all PoW based blockchains is still an essential issue. Generally speaking, all well-designed difficulty control algorithms should meet several demands, which has explicitly announced by Ethereum \cite{designrationale} and we concentrate them into two:

1. Simplicity \& low memory: The algorithm should not rely on too much blocks of history and easy to implement. In order to reach a consensus, the block difficulty can only be calculated according to the information in the block headers.

2. Fast updating \& low volatility: The block produce time should be able to be readjusted quickly along with the changes of network hash rate and the block difficulty should not bounce excessively if the network hash rate is constant. What is more, the difficulty control algorithm should not excessively encourage miners to fiddle with timestamps.

%The difficulty of one block can be three times higher than the difficulty of its parent block if the block time is one and a third times lower than expected. The difficulty control algorithm for Verge has an excessive incentive for miners to manipulate the block difficulty. Thus, the attacker only spent a little to change the difficulty of Verge to zero within several blocks by submitting blocks with fake timestamps.

In this paper, we propose a general difficulty control algorithm which is applicable to all PoW based blockchains and provide a typical example for Ethereum which meets the demands better. Meanwhile, the proposed method is also able to conduct the anomaly detection based on the analysis by a two-layer neural network.

%Emerging blockchains are striving to increase the volume of the block and shorten the block time, thereby reducing the time to verify each transaction, which is more attractive to users. Ethereum chooses the block time to be approximately 12 seconds and a block time shorter than 12 seconds will increase the orphan rate of blocks, causing the chain harder to converge \cite{buterin2014toward}.

\section{Problem Formulation}
\label{Problem Formulation}
Once one block $B_k$ ($k$ is the height of the block) is generated by a miner, the miner will set a timestamp $t_k$ for $B_k$ and then broadcast $B_k$ into the network. The initial timestamp $t_0 = 0$ and $t_0 \leq t_1 \leq t_2 \leq \cdots$. The time between two continuous blocks is denoted with $T_k = t_k-t_{k-1}$, which is called the block produce time for block $B_k$.

The block produce time $T_k$, which is exponentially distributed, depends on the block difficulty, the real-time network hash rate and other delays. The expectation of $T_k$ is $\tau_k$, given by

\vspace{-5pt}
\begin{equation}
	\tau_k = \frac{D_k}{h_k}+t^p_k =  \frac{D_k}{H_k},
\end{equation}

\noindent
where $D_k$ is the block difficulty, $h_k$ is the actual real-time network hash rate and $t_k^p$ is the overall delay which includes the propagation delay among peers and the delay for peers to verify the block. The $t_k^p$ and $h_k$ are unavailable. Thus we propose the nominal hash rate $H_k$, which provides more convenience for analysis. 

We estimate the nominal hash rate $H_k$ for each block by dividing the average block difficulty with the average block produce time, where $W$ is the length of the sliding window.

%\begin{equation}
%	H_k^e =\frac{D_k M}{\sum_{i=1}^M T_{k-M+i}} 
%\end{equation}

\vspace{-5pt}

\begin{equation}
	H_k =\frac{\sum_{i=1}^W D_{k-W+i}}{\sum_{i=1}^W T_{k-W+i}}.
\end{equation}

\subsection{Nominal Hash Rate Change from Data}
\label{section2.1}
Suppose that the nominal hash rate changes periodically and the period is represented by the number of blocks and is also $W$. In this view, the periodically changing nominal hash rate $H^*$ is given by

\vspace{-17pt}
\begin{equation}
	H_{nW+1}^*=H_{nW+2}^*=\cdots=H_{(n+1)W}^*=\frac{\sum_{i=1}^W H_{nW+i}}{W}.
\end{equation}

\begin{figure}[tbp]
\begin{center}
   \includegraphics[width=0.95\linewidth]{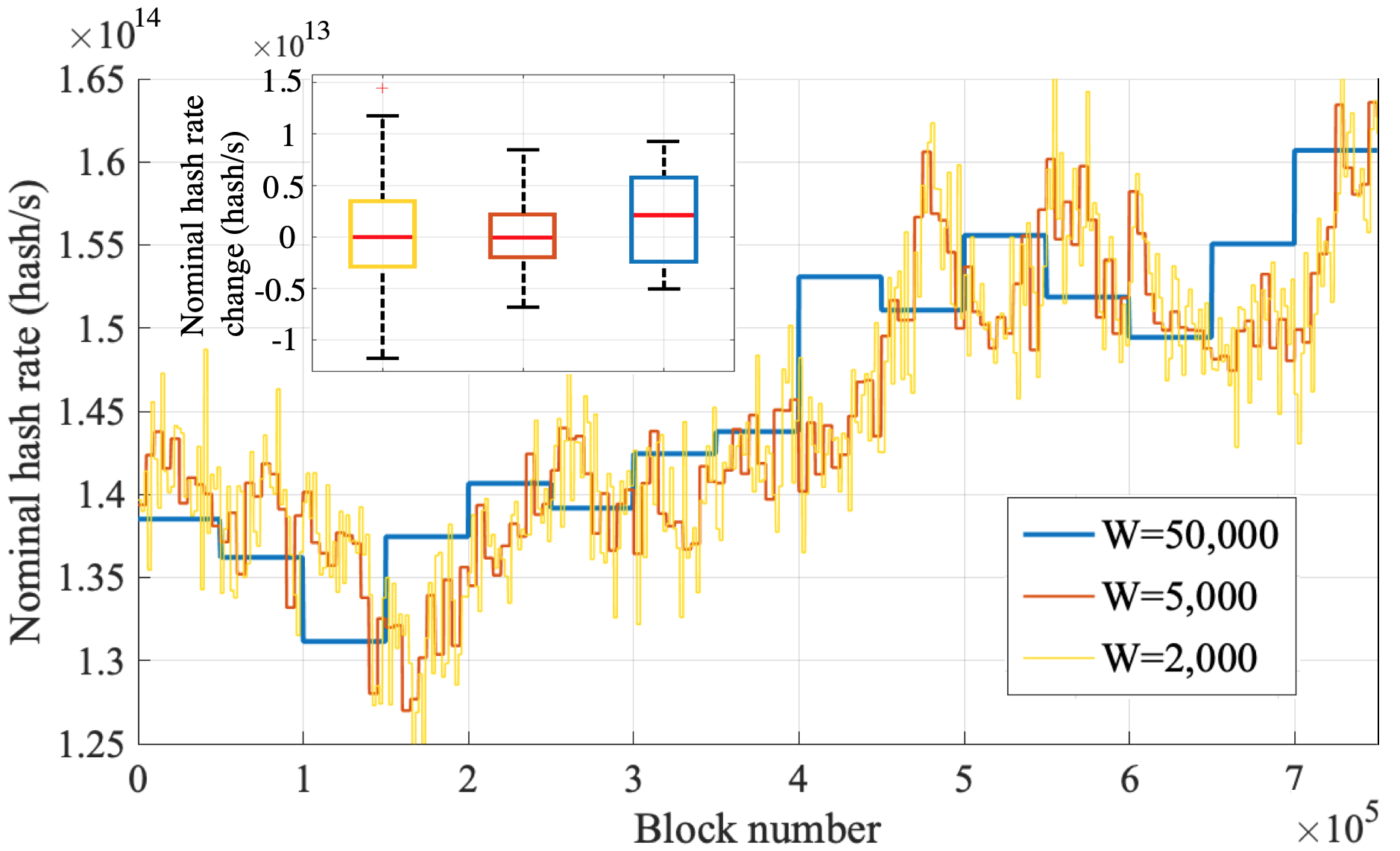}
\end{center}
\vspace{-15pt}
   \caption{The periodically changing nominal hash rate and the boxplots of the nominal hash rate change with different periods.}
   \vspace{-15pt}
\label{distribution}
\end{figure}

\noindent
We collect the block produce time and the block difficulty of continuous 800,000 blocks in the Constantinople stage of Ethereum, and then calculate the periodically changing nominal hash rate. The periodically changing nominal hash rate for Ethereum and the boxplots of nominal hash rate change $\Delta H_n^*= H_{nW+1}^*-H_{nW}^*$ with different $W$ are shown in Fig. \ref{distribution}. As revealed in the boxplots in Fig. \ref{distribution}, the nominal hash rate change is numerically normally distributed and appears much randomness when $W=2,000$ and $5,000$. When $W=50,000$, the mean of the nominal hash rate change is not zero, then an increasing trend of nominal hash rate begins to appear.

\subsection{A General Difficulty Control Algorithm}
A general difficulty control algorithm is written as

\vspace{-7pt}
\begin{equation}
	D_k = D_{k-1}-D_{k-1}\cdot I_k\cdot f(T_{\textbf{previous}}),
\end{equation}
\vspace{-15pt}

\noindent
where $D_{k-1}$ is the difficulty of the last block, $I_k$ is the indicator and $f(T_{\textbf{previous}})$ is the update function, in which $T_{\textbf{previous}}$ is the summation of continuous previous block produce times. As a special case, the difficulty control algorithm for Ethereum satisfies the following condition:

\vspace{-12pt}

\begin{equation}
\begin{cases}
I_k = 1, \forall k\\
T_{\textbf{previous}} = T_k\\
f(T_{\textbf{previous}}) = 
\begin{cases}
{(\lfloor \frac{T_{\textbf{previous}}}{9} \rfloor -1)/2048}, & 0< T_k\leq 900\\
99/2048, & T_k> 900.
\end{cases}
\end{cases}
\label{1}
\end{equation}
\vspace{-8pt}

\noindent
%For Ethereum, there is only one element $T_k$ in $T_{\textbf{previous}}$. 
The difficulty control algorithm for Bitcoin satisfies the following conditions:

\vspace{-8pt}

\begin{equation}
\begin{cases}
I_k = 
\begin{cases}
1, \text{  if } k \text{ mod } N = 0\\
0, \text{  otherwise}
\end{cases}\\
T_{\textbf{previous}} = \sum_{i=1}^N T_{k-i+1}\\
f(T_{\textbf{previous}}) = 1-N\beta/{T_{\textbf{previous}}},
\end{cases}
\end{equation}
\vspace{-5pt}

\noindent
where $N=2016$ and $\beta = 10$ minutes, which is the expected block produce time. Then the $T_{\textbf{previous}}$ for Bitcoin is the summation of 2016 previous block produce times.

\section{Algorithm Design}
\label{scheme}

\subsection{Design of the indicator}
The existence of the indicator $I_k$ is to control the updating speed of the algorithm. When $I_k$ is approaching 0, the difficulty change is suppressed. It is necessary to recognize different patterns of the block produce time change and react to the cases behind those patterns properly.

\begin{figure}[htbp]
\begin{center}
   \includegraphics[width=0.9\linewidth]{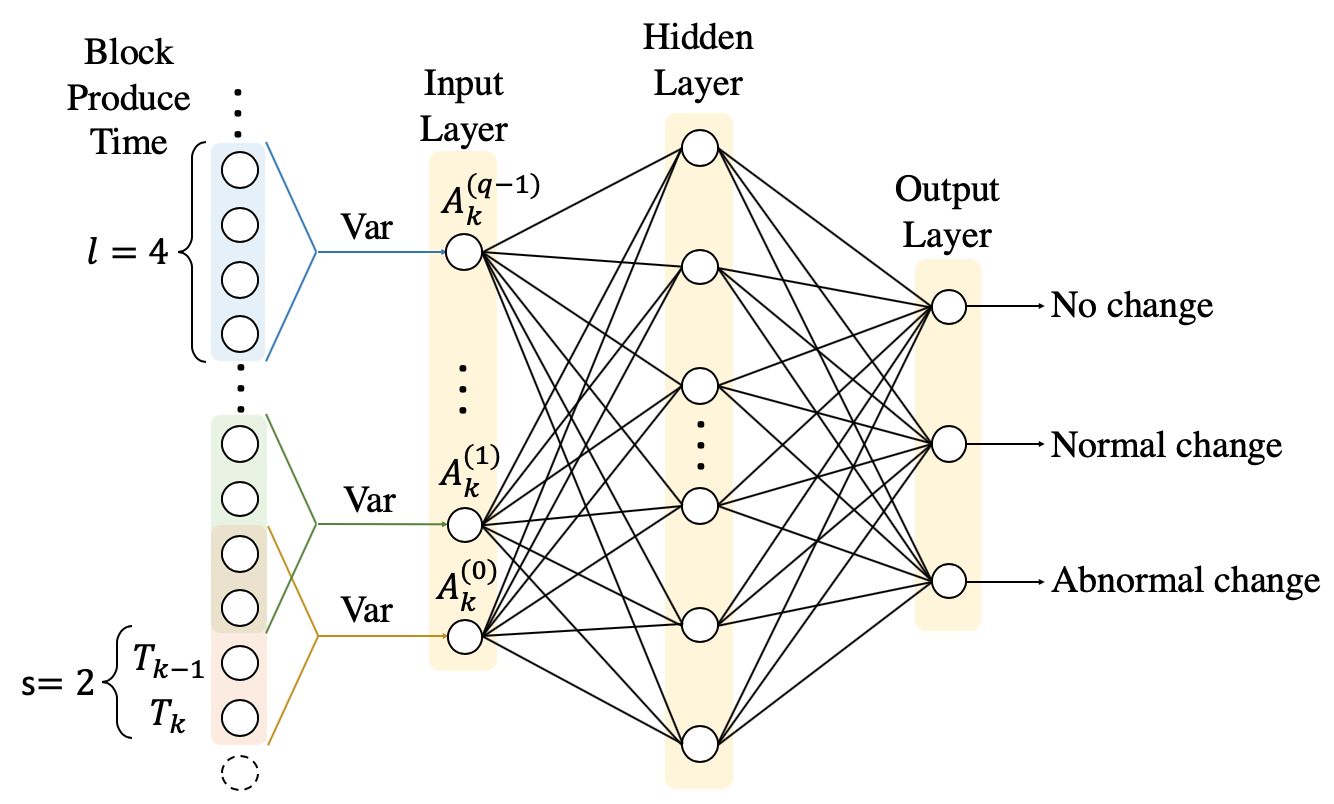}
\end{center}
\vspace{-15pt}
   \caption{The illustration of the two-layer neural network.}
\label{network}
\end{figure}

There are three patterns of the trend of the block produce time: no change (case 1), normal change (case 2) and abnormal change (case 3). Different combinations of previous block produce times can be used as input features to estimate the state of the blockchain. The two-layer neural network shown in Fig. \ref{network} is able to recognize different patterns by distinguishing the trend of the variance of block produce time.

The probabilities of those three patterns are defined as $P^{(2)}_k, P^{(2)}_k$ and $P^{(3)}_k$ separately. We set $I_k=P^{(2)}_k$, which enables the algorithm to change the difficulty normally only when the neural network recognizes the change of the block produce time is normal.

There are $q$ inputs of the neural network which are $A_k^{(0)}, A_k^{(1)}, \cdots, A_k^{(q-1)}$. The input $A_k^{(j)}$ uses $s$ blocks ahead of $A_k^{(j+1)}$. The number of block produce times used to calculate the variance as an input feature is $l$. The expression of $A_k^{(j)}$ is written as

\vspace{-15pt}
\begin{equation}
\begin{split}
	A_k^{(j)}&=\mathbb{E}\left((T^{(j)}_k)^2\right)-\left(\mathbb{E}(T^{(j)}_k)\right)^2\\
	&=\frac{1}{l}\sum_{i=k-js-l+1}^{k-js} T_i^2-\left(\frac{1}{l}\sum_{i=k-js-l+1}^{k-js} T_i\right)^2 \\
	&\qquad (j=0,1,\cdots, q-1).
\end{split}
\end{equation}
\vspace{-5pt}

\noindent
It seems that calculating $A_k^{(j)}$ requires $l$ block produce times and $q$ features requires $(q-1)s+l$ block produce times in total, which is very memory-cost. But in fact, the updating from $A_{k-1}^{(j)}$ to $A_k^{(j)}$ only requires a little more information. By reducing $q$, the required memory can be reduced further. Moreover, if block difficulty is only changed exactly every $s$ blocks, then the neural network can fully use the existed knowledge by calculating $A_k^{(j)}=A_{k-s}^{(j-1)} (j=1,2,\cdots, q-1)$ and only the input $A_k^{(0)}$ needs a totally update.

Even though complicated deep neural networks can distinguish the patterns more accurately, we choose a rather simple neural network with only one hidden layer out of the simplicity demand and the fast calculation demand.

\subsection{Design of the Update Function}

\begin{figure}[tbp]
\begin{center}
   \includegraphics[width=1\linewidth]{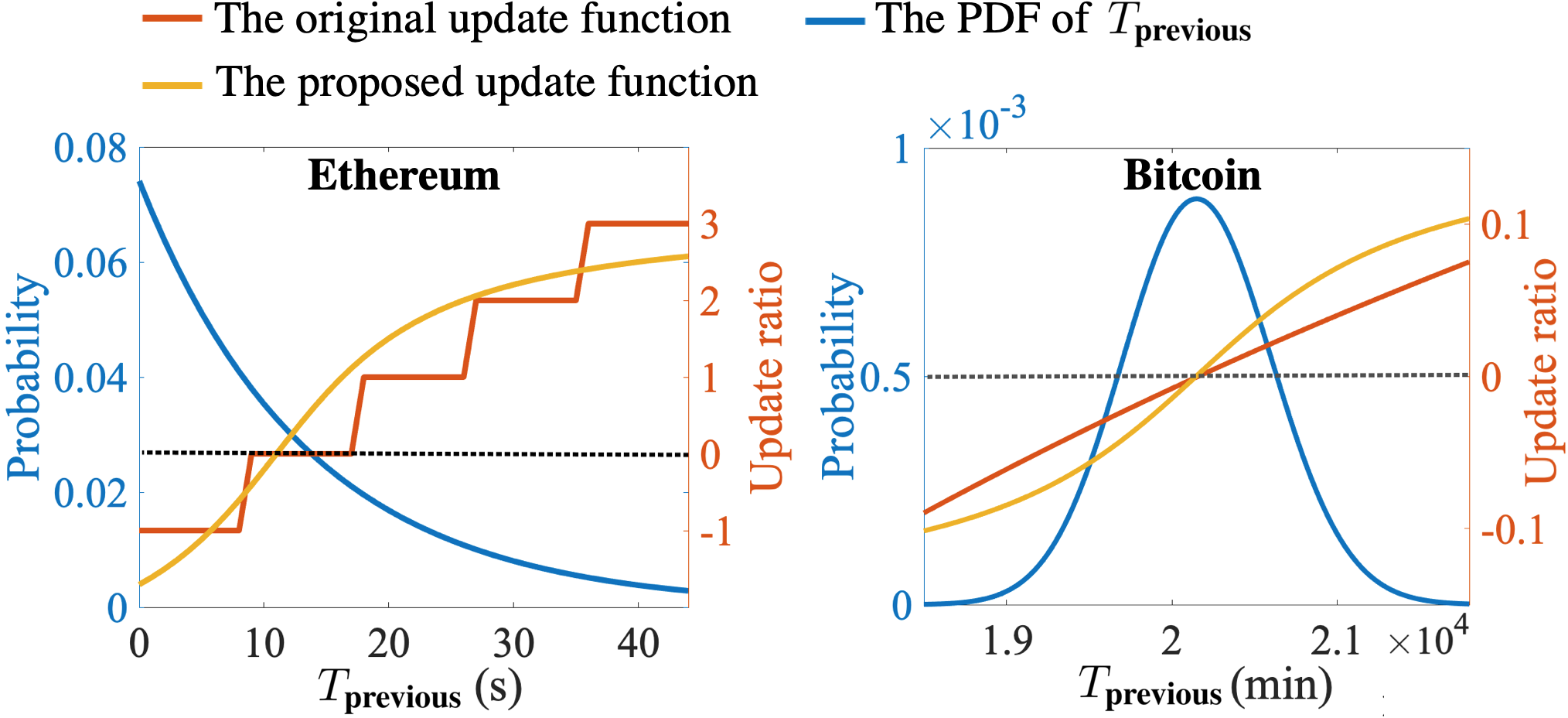}
\end{center}
	\vspace{-15pt}
   \caption{The update functions and the PDF of $T_{\textbf{previous}}$ of Ethereum (left) and Bitcoin (right).}
   \vspace{-10pt}
\label{update_function}
\end{figure}

The following condition should be satisfied to promise the block produce time to converge to the expected value.

\textit{Condition 1:}
\begin{equation}
\int_{0}^\infty f(T_{\textbf{previous}})\cdot g(T_{\textbf{previous}}) \text{d}T_{\textbf{previous}} = 0,
\end{equation}

\noindent
where $g(T_{\textbf{previous}})$ is the probability density function of $T_{\textbf{previous}}$. If $T_{\textbf{previous}}=T_k$ which is the case of Ethereum, then $g(T_{\textbf{previous}}) = g(T_k) = \frac{1}{\beta} e^{-\frac{T_k}{\beta}}$ and $\beta$ is the expected block produce time which will converge to. It is easy to prove and test that the $f( T_{\textbf{previous}})$ of Ethereum in Equation (\ref{1}) can promise an expected 13.5 seconds block produce time. In the case of Bitcoin, $T_{\textbf{previous}}$ is the summation of multiple exponentially distributed random variables, so $g(T_{\textbf{previous}})$ is the probability density function of an Erlang distribution.

The blockchain network is hardly possible to generate blocks with continuous long block produce times. Continuous long block produce times will change the block difficulty vastly and quickly, thereby challenging the robustness of the blockchain. Both Bitcoin and Ethereum handle this by setting up an additional bound for the change of block difficulty. A proper update function can avoid this redundant design. The $\arctan$ function has an upper bound and a lower bound inherently. Thus, a shifted and scaled $\arctan$ function is a feasible update function that satisfies Condition 1, whose general form is

\vspace{-15pt}
\begin{equation}
	f(T_{\textbf{previous}})=A(\arctan(B(T_{\textbf{previous}}-C))+D).
\end{equation}
\vspace{-12pt}

\noindent
The parameter $A$ determines the volatility of the block difficulty, while proper $B, C$ and $D$ together guarantee Condition 1. Feasible sets of parameters for Ethereum and Bitcoin are shown in Table \ref{table0}.

\begin{table}[h]
\newcommand{\tabincell}[2]{\begin{tabular}{@{}#1@{}}#2\end{tabular}}
\caption{Parameters in the proposed update functions for Ethereum and Bitcoin}
\vspace{-15pt}
\begin{center}
\begin{tabular}{|l|c|c|c|c|}
\hline
Blockchain & A & B & C & D\\
\hline
Ethereum & 1e-3 & 1e-2 & 11 seconds & 0\\
\hline
Bitcoin & 5e-5 & 1e-3 & 20,160 minutes & 1.35e-3\\

\hline
\end{tabular}
\end{center}
\vspace{-15pt}
\label{table0}
\end{table}

%\begin{equation}
%	f_1(\sum T_{\textbf{previous}})=f_1(T_k)=\frac{1}{\mu}\times
%	\begin{cases}
%	-1, &T_k\leq 9\\
%	0, &9<T_k\leq 30\\
%	4.4, &30<T_k\leq 40\\
%     8.8, &40<T_k\leq 50\\
%     2.2, &T_k>50\\
%	\end{cases}
%\end{equation}

%where $\mu$ is 2048, the same as the parameter in the original update function for Ethereum. 
%As shown in Fig. \ref{update_function}, the difficulty changes slightly using the new update function if block time is long. $\sum T_{\textbf{previous}}$ with low possibility should induce small change in difficulty because it may demonstrate abnormal cases. $\sum T_{\textbf{previous}}$ with high possibility should also induce small change in difficulty because its appearance is normal.

%If $T_{\textbf{previous}}$ includes more than one block time, a possible update function $f_2(\sum T_{\textbf{previous}})$ is shown in Fig. \ref{update_function_2}. In this example, $L=10$.

%\begin{equation}
%	f_2(\sum T_{\textbf{previous}})=f_2(\sum_{i=0}^L T_{k-i})=\frac{1}{\mu}\times
%	\begin{cases}
%	-1, &T_k\leq 13\\
%     1, &T_k>13\\
%	\end{cases}
%\end{equation}

The curves of the proposed difficulty update functions and the original difficulty update functions for Ethereum and Bitcoin are shown in Fig. \ref{update_function}.

\section{Simulations}
\label{Simulation}

\begin{figure*}[htbp]
\begin{center}
   \includegraphics[width=1\linewidth]{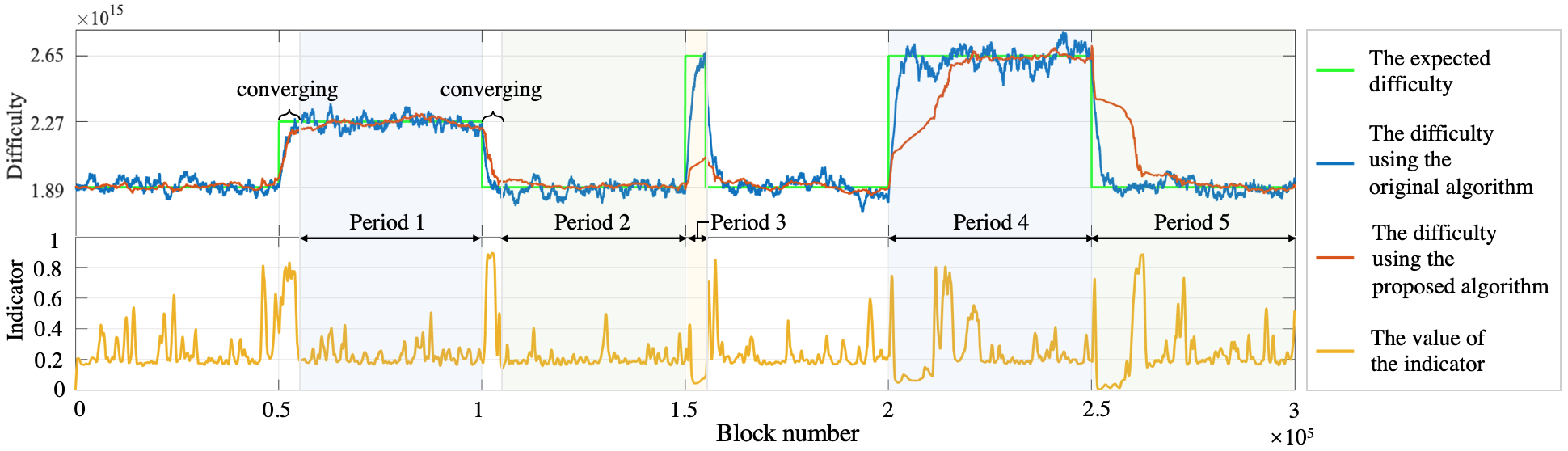}
\end{center}
  \vspace*{-15pt}
   \caption{The block difficulty and the indicator in the simulation.} 
    \vspace*{-10pt}
\label{simu}
\end{figure*}

We simulate the change of the nominal hash rate based on real data from Ethereum and then compare the effects of different difficulty control algorithms.

As explained in Section \ref{section2.1}, much randomness is debilitated and the trend of hash rate appears when $W=50,000$, so $W=50,000$ is a reasonable period. In real historical data, $\Delta H_n^*$ is within $\pm 1e13$ hash/s as shown in Fig. \ref{distribution}. There exists a range of the normal hash rate change. For Ethereum, we suppose that the hash rate change should not exceed $\mathcal{B} = \pm 20\%$ of the original average hash rate. A sudden network hash rate change more than $\pm 20\%$ of the original average hash rate is considered as an anomaly.

%(800000 Ethereum blocks recorded from 2019-03-04 to 2019-07-06)

\subsection{The Neural Network}
We conduct Monte-Carlo simulation in the training process. Each sample begins with a hash rate of $1.455e14$ hash/s. The hash rate change for each sample ranges from $-60\%$ to $+60\%$ of the beginning hash rate, including the cases of normal change and abnormal change. The sudden change happens between $B_c$ and $B_{c+1}$. Thus, the block produce time sequence begins with $T_{c-s(q-1)-l}$ and ended with $T_{c-1}$ is the input whose output is labelled as case 1. The block produce time sequence begins with $T_{c-l+1}$ and ended with $T_{c+s(q-1)}$ is the input whose output is labelled as case 2 or case 3. If the hash rate change is within $\mathcal{B} = \pm 20\%$ of the beginning hash rate, then the sample is case 2, otherwise case 3. The amounts of these three cases for training are equal.

The accuracy of the neural network mainly depends on the production of $s$ and $q-1$, which is the number of blocks after the change happens. In other words, the accuracy of the neural network will increase steadily as time passes after the sudden hash rate change. Specifically, the overall accuracy of the neural network is $78.41\%$ after 1,000 blocks (3.75 hours) from the change and $88.84\%$ after 5,000 blocks (18.75 hours) from the change. 

%In the analysis below, the classifier we use is trained by given 2000 blocks-later data.
For the neural network in our experiments, we set $s=200, q=11$ and $l=2,000$. The number of neurons in the hidden layer is 25.

%In practice, the number of input features is $q=11$, the number of blocks between two features $s=200$ and the number of block times used to calculate the variance as an input featur%e $l=2000$. The relationship between $s, l$ and the accuracy of the network is shown in

%\begin{figure}[tbp]
%\begin{center}
%   \includegraphics[width=0.7\linewidth]{Acc.png}
%\end{center}
%   \caption{The accuracy of case 2 and the overall accuracy vs. $s$ and the number of input features.}
%\label{Acc}
%\end{figure}

\subsection{Performance of Difficulty Control Algorithms}
As shown in Fig. \ref{simu}, an additional $20\%$ Ethereum's hash rate is injected to the mining pool at the 50,000th block and withdrew at the 100,000th block in the simulation. An additional $40\%$ Ethereum's hash rate is injected to the mining pool at the 150,000th block, the 200,000th block and withdrew at the 155,000th block, the 250,000th block. The mean and the variance of the block difficulty in Period 1 and Period 2 as shown in Fig. \ref{simu} are calculated to evaluate the performance of different difficulty control algorithms. As shown in Fig. \ref{simu}, the convergence time after the sudden change using two algorithms are similar, but it is obvious that the proposed algorithm can provide a smoother block difficulty. The numerical simulation result is shown in Table. \ref{table1}.

\begin{table}[h]
\newcommand{\tabincell}[2]{\begin{tabular}{@{}#1@{}}#2\end{tabular}}
\caption{Properties of the block difficulty using different difficulty control algorithms in Period 1 and Period 2.}
  \vspace{-10pt}
\begin{center}
\begin{tabular}{|l|c|c|c|c|}
\hline
Property & Original & Proposed & Reduced by\\
\hline \hline
MSE (Period 1) & 2.048e27 & 6.929e26 & 66.3\%\\
\hline
Mean (Period 1) & 2.255e15 & 2.276e15 & /\\
\hline
MSE (Period 2) & 1.632e27 & 3.829e26 & 76.7\%\\
\hline
Mean (Period 2) & 1.885e15 & 1.891e15 & /\\

\hline
\end{tabular}
\end{center}
  \vspace{-10pt}
\label{table1}
\end{table}

When an abnormal case occurs, the proposed algorithm will suppress the changing of block difficulty until it has passed long enough to assure the correctness of the injection or withdrawal of the hash rate, which is illustrated as Period 4 and Period 5 in Fig. \ref{simu}. If an abnormal increment or decrement of hash rate only lasts for a short period of time which is more likely to be a malicious attack, which is illustrated as Period 3 in Fig. \ref{simu}, the proposed algorithm is tend to pass this period smoothly.

%We simulate an abnormal hash rate injection and then a equal withdraw to watch how different difficulty algorithms will respond to it. In the simulation, the abnormal injected hash rate is 40\% and 60\% of the original hash rate. As the experiment result shown in Fig. \ref{simu}, $P_k^3$ for different injections are totally disparate and the 40\% and 60\% injections, as abnormal cases, both cause continuous high probability of $P_k^3$.

%As shown in Fig. \ref{simu}, as soon as the abnormal hashrate injection is detected, the new algorithm will begin to suppress the changing of the difficulty. 
Once client security bugs or other black-swan issues happen and cause the block produce time happen to be very long continuously, the difficulty control algorithms are supposed to keep the block difficulty unchanged. The amplitude of the block difficulty change in this case depends on the bound of the update function. The amplitude of the block difficulty change of the original algorithm is $\frac{\mathrm{max}(f_{\mathrm{ori}}(T_{\textbf{previous}}))}{\mathrm{max}(f_{\mathrm{new}}(T_{\textbf{previous}}))}=\frac{99/2048}{1e-3\lim_{t\rightarrow \infty} \arctan (t)}\approx \frac{99}{\pi}$ times of the proposed algorithm's.

\section{Conclusion}
\label{conclusion}
In this paper, we propose a general difficulty control algorithm for PoW based blockchains and also provide an alternative algorithm for Ethereum. Simulations based on real data reveal that our neural network-based algorithm preserves the fast updating as well as the low volatility of the block difficulty. The proposed algorithm is able to detect anomaly and handle abnormal cases properly.

\bibliographystyle{IEEEtran}
\bibliography{ref}

\end{document}